\documentclass[a4paper]{PoS}

\title{Pulsar emission in the very-high-energy regime}

\ShortTitle{Pulsar emission in the very-high-energy regime}

\author{\speaker{M. Breed},$^{a}$ C. Venter$^{a}$ and A. K. Harding$^{b}$\\
\llap{$^a$}Centre for Space Research, North-West University, Potchefstroom Campus, Private Bag X6001,\\ Potchefstroom 2520, South Africa\\
\llap{$^b$}Astrophysics Science Division, NASA Goddard Space Flight Center, Greenbelt, MD 20771, USA\\
E-mail: \email{monicabarnard77@ gmail.com}}

\abstract{The vast majority of the pulsars detected by the \textit{Fermi} Large Area Telescope (LAT) display spectra with exponential cutoffs falling in a narrow range around a few GeV. Early spectral modelling predicted spectral cutoff energies of up to 100~GeV. More modern studies estimated spectral cutoff energies in the 1$-$20~GeV range. It was therefore not expected that pulsars would be visible in the very-high-energy (VHE; >100~GeV) regime. The \textit{VERITAS} detection (confirmed by \textit{MAGIC}) of pulsed emission from the Crab pulsar up to 400~GeV (and now possibly up to 1~TeV) therefore raised important questions about our understanding of the electrodynamics and local environment of pulsars. \textit{H.E.S.S.} has now detected pulsed emission from the Vela pulsar in the 20$-$120~GeV range, making this the second pulsar detected by a ground-based Cherenkov telescope. We will review the latest developments in VHE pulsar science, including an overview of recent observations and refinements to radiation models and magnetic field structures. This will assist us in interpreting the VHE emission detected from the Crab and Vela pulsars, and predicting the level of VHE emission expected from other pulsars, which will be very important for the upcoming \textit{CTA}.}

\FullConference{3rd Annual Conference on High Energy Astrophysics in Southern Africa - HEASA2015,\\
		18-20 June 2015\\
		University of Johannesburg, Auckland Park, South Africa}

\begin{document}


\section{Introduction}\label{sec:history}

Since the successful launch of the {\it Fermi} LAT \cite{Atwood2009}, a high-energy (HE) satellite measuring $\gamma$-rays in the range 20~MeV$-$300~GeV, two pulsar catalogues (1PC, \cite{Abdo2010b}; 2PC, \cite{Abdo2013}) have been released, collectively discussing the light curve and spectral properties of 117 pulsars. The vast majority of the {\it Fermi}-detected pulsars display exponentially cutoff spectra with cutoffs around a few GeV. These spectral cutoffs are believed to be a signature of curvature radiation (CR), which is assumed to be the dominating radiation process in the GeV band.

There exist several radiation models that can be used to study HE emission from pulsars. These include the polar cap (PC; \cite{Daugherty1982}), slot gap (SG; \cite{Arons1983}), outer gap (OG; \cite{Cheng1986b}), and the pair-starved polar cap (PSPC; \cite{Harding2005}) models. These models make different assumptions for the geometry and location of their `gap regions' (where particle acceleration takes place due to an unscreened, rotation-induced $E$-field parallel to the local $B$-field, as well as subsequent emission by these particles). To predict realistic HE emission in these standard pulsar models, one has to take detailed particle transport and radiation mechanisms into account. These mechanisms include CR, synchrotron radiation (SR), and inverse Compton scattering (ICS). More recently, pulsar wind models were developed that postulate that HE emission originates beyond the light cylinder. These include the striped wind model in which pulsed HE emission arises from SR by relativistically hot particles that are accelerated via magnetic reconnection inside the current sheet \cite{Petri2011}.         

Early modelling, assuming the standard OG model, predicted spectral components in the VHE regime when estimating the ICS flux of primary electrons on SR or soft photons (\cite{Cheng1986b,Romani1996,Hirotani2001}). This resulted in a natural bump around a few TeV (involving $\sim$10~TeV particles) in the extreme Klein-Nishina limit. However, these components may not survive up to the light cylinder (where the pulsar's corotation speed equals the speed of light) and beyond, since two-photon pair production may lead to absorption of the TeV $\gamma$-ray flux \cite{Hirotani2001}. 

Later studies assumed standard pulsar models and CR to be the dominant radiation mechanism producing HE $\gamma$-ray emission when performing spectral modelling and found spectral cutoffs of up to 100~GeV. For example, \cite{Bulik2000} modelled the cutoffs of millisecond pulsars (MSPs), which possess relatively low $B$-fields and short periods. Their model assumed a static dipole $B$-field, a PC  geometry (emission from HE particles originates close to the stellar surface and the particles accelerate up to a few stellar radii), and a classical $E$-field expression, also including one-photon (magnetic) pair creation. Their estimated spectrum cut off at $\sim$100~GeV. A similar study was done by \cite{Harding2002} who assumed a general relativistic (GR) corrected $E$-field. They also found CR spectral cutoffs at energies between 50$-$100~GeV. The X-ray and $\gamma$-ray spectra of rotation-powered MSPs were investigated by \cite{Harding2005} using a PSPC model (i.e., emission starting at the stellar surface up to the light cylinder over the full open volume) and a GR-corrected $E$-field, and found CR cutoffs of $\sim$10$-$50~GeV (see also \cite{Venter2005,Frackowiak2005}). Therefore, they concluded that the HE CR from MSPs occurred in an energy band that was above the detection range of satellite detectors like EGRET and below that of ground-based Cherenkov detectors. The optical to $\gamma$-ray emission spectrum was modelled by \cite{Harding2008} for the Crab pulsar assuming an SG accelerator (i.e., radiation from narrow gaps starting at the stellar surface up to the light cylinder) and a retarded vacuum dipole (RVD) $B$-field, and found spectral cutoffs of up to a few GeV. Phase-resolved spectra of the Crab pulsar were also modelled by \cite{Hirotani2008} using the OG and SG models, predicting HE cutoffs of up to 25~GeV. Another study used the RVD $B$-field, an extended OG model and a synchrotron-self-Compton (SSC) radiation mechanism (where relativistic particles upscatter the SR photons emitted by the same particle population) to predict phase-resolved spectra for the Crab, and found HE cutoffs around $\sim$10~GeV \cite{Tang2008}.
 
In view of the above theoretical paradigm it was not expected that pulsars should be visible in the VHE regime. Later observations therefore challenged this standard paradigm. In this paper we review the latest VHE observations (Section~\ref{sec:observations}) and theoretical developments (Section~\ref{sec:theories}) which may help explain these observations. Our conclusions follow in Section~\ref{sec:conclusions}.

\begin{figure}[t]
\begin{minipage}{9cm}
	\includegraphics[width=9cm]{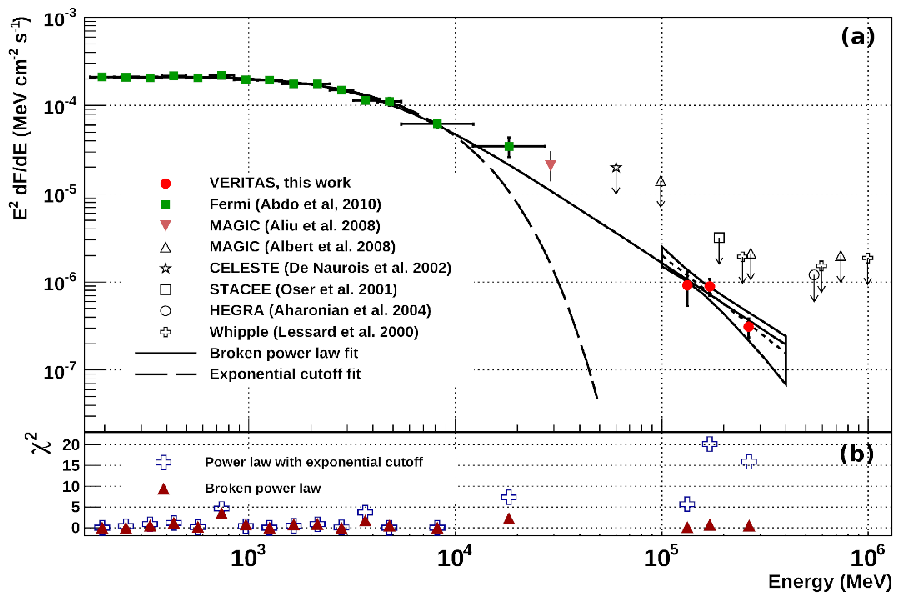}
\end{minipage}
\hspace{0.2cm}
\begin{minipage}{6.3cm}\vspace{0.1cm}
	\includegraphics[width=6.3cm]{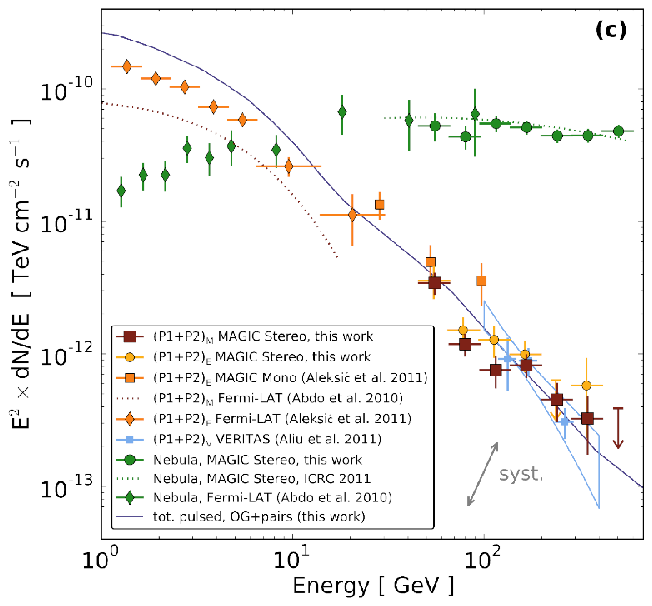}
\end{minipage}
\caption{\label{fig:VMspectra} The observed and modelled spectra of the $\gamma$-ray emission from the Crab pulsar (including both light curve peaks P1 and P2). Panel (a) represents the data obtained by \textit{VERITAS} (red solid circles) and also includes measurements by \textit{Fermi} LAT (green squares), the measurement >25~GeV from \textit{MAGIC} (filled pink triangle), as well as upper limits from several other telescopes (empty markers). The dashed line specifies an exponentially cutoff  power law fit, whereas the solid line represents a broken power law fit. From panel (a) and (b) it is clear that the \textit{VERITAS} observations significantly prefer a broken power law above an exponentially cutoff power law \cite{Aliu2011}. In panel (c) the phase-resolved emission spectrum as measured by \textit{MAGIC} (dark red squares) from the Crab >400~GeV is shown, together with an OG model that includes emission from pairs (solid line; \cite{Aleksic2012}).}
\end{figure}

\section{Observational revolution}\label{sec:observations}

\begin{figure}[t]
\centering
\includegraphics[width=10cm]{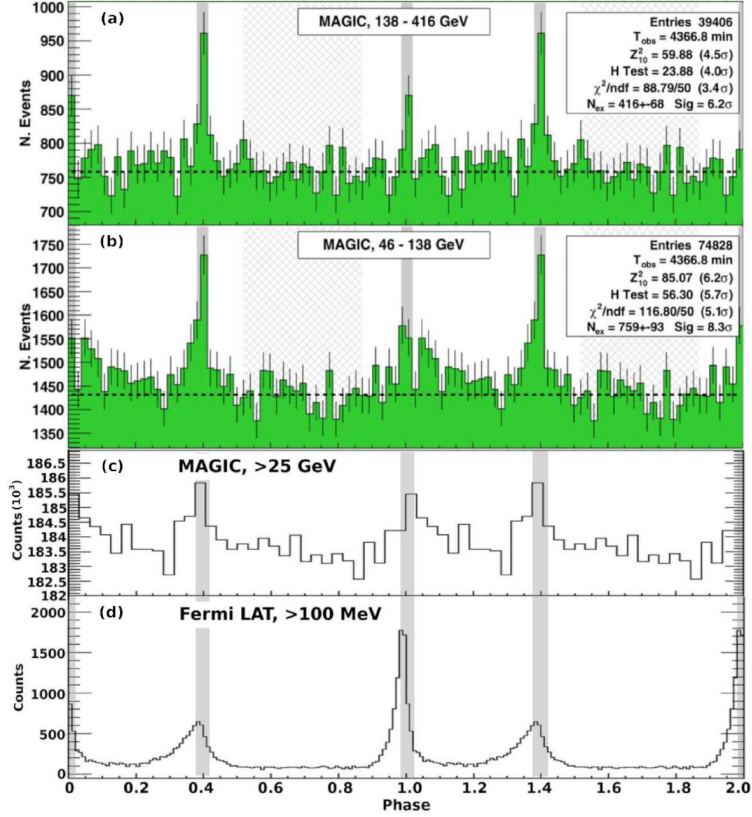}
\caption{\label{fig:MAGIC_LC} Pulse profiles of the Crab pulsar in different energy ranges, increasing from bottom to top, as obtained by \textit{MAGIC} (panels (a)-(c), \cite{Aleksic2012}) and \textit{Fermi} (panel (d), \cite{Abdo2010a}). Three effects are visible in the energy-dependent profiles: the peaks remain at the same phases (shaded light grey areas), the P1/P2 ratio decreases as energy increases (at energies >25~GeV the peaks are nearly equal in height), and the pulse width decreases with increasing energy. The dashed line is the background level.}
\end{figure}
In view of existing HE spectral model predictions, detection of pulsar emission in the VHE regime was not expected. \textit{MAGIC} observed the Crab pulsar and detected pulsed emission above $\sim25$~GeV (\cite{Aliu2008}, see pink triangle in Figure~\ref{fig:VMspectra}(a)). This raised the important question of whether the single CR component extends to higher energies than previously thought. It was very surprising when \textit{VERITAS} announced the detection of pulsed emission from the Crab pulsar above $\sim100$~GeV \cite{Aliu2011}. In Figure~\ref{fig:VMspectra}(a) and (b) the \textit{VERITAS} data are best described by a broken power law. \textit{MAGIC} confirmed this spectral shape (see Figure~\ref{fig:VMspectra}(c)) when they observed emission $\sim$400~GeV~\cite{Aleksic2012}, soon after their initial detection of emission >25~GeV. This implied severe constraints on the spectral interpretation: invoking only a single CR component (or e.g., IC), or more than one (e.g., CR and SSC) qualitatively new spectral components.

Three trends were noted in the energy-dependent pulse profiles. First, the peaks remain at the same phase $\phi$, i.e., the main pulse P1 at $\phi=[-0.1,0.13]$ and the interpulse P2 at $\phi=[0.35,0.45]$. Second, the ratio of intensity of P1 vs. that of P2 (i.e., P1/P2) decreases as energy increases. At energies >25~GeV, ${\rm P1/P2}\approx1$. Lastly, the pulse width decreases with an increase in energy. These effects are illustrated in Figure~\ref{fig:MAGIC_LC} which shows pulse profiles of the Crab pulsar in different energy ranges (increasing from bottom to top panel) as obtained by \textit{MAGIC} (panels (a)-(c)) and \textit{Fermi} (panel (d)). \textit{MAGIC} detected significant bridge emission above 50~GeV between $\phi=[0.024,0.37]$ \cite{Aleksic2012,Saito2014}, consistent with the {\it Fermi} observations \cite{Abdo2010a}. The intensity and profile shape of the bridge emission vary as a function of energy. At energies <10~GeV the spectrum for the bridge emission (emission between P1 and P2) is notably harder than that of P1 and P2. \textit{MAGIC} has just claimed detection of photons with energies up to 1~TeV coinciding with P2 in the Crab's profile (Aleksic et al. 2015, in prep.). The detection of emission >400~GeV has not been confirmed by \textit{VERITAS} to date \cite{Nguyen2015}. However, they are continuing to search for VHE pulsed emission in their archival data of 19 known pulsars with $\sim$20 hours of observations each \cite{Archer2015}. 

\begin{figure}[t]
\centering
\includegraphics[width=10cm]{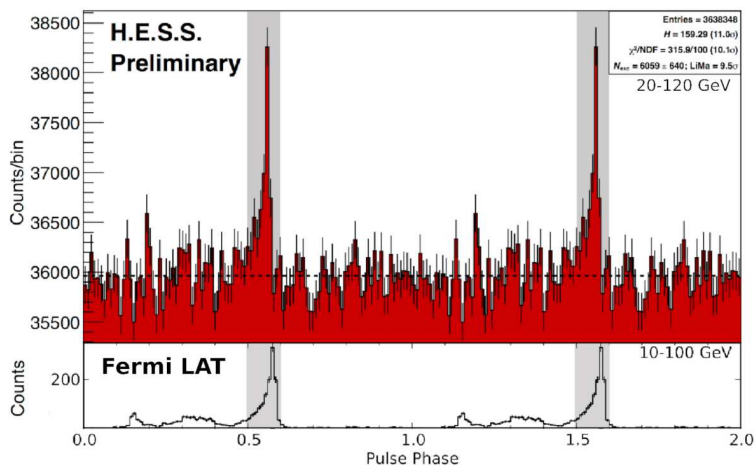}
\caption{\label{fig:HF_LC} Light curves obtained by \textit{H.E.S.S.} in 20$-$120~GeV range (top panel, \cite{Stegmann2014}) and \textit{Fermi} in 10$-$100~GeV range (bottom panel, \cite{Leung2014}). From the vertical shaded regions on each light curve it is evident that the P2 peaks are phase-aligned.}
\end{figure}

\begin{figure}[t]
\centering
\includegraphics[width=10cm]{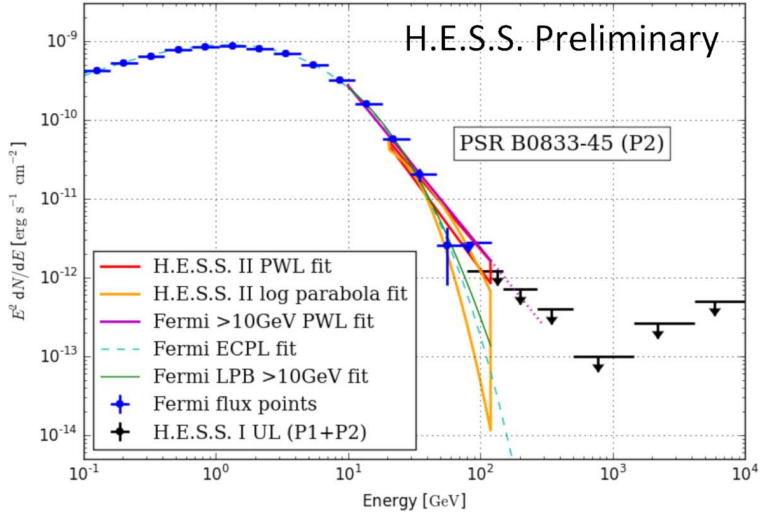}
\caption{\label{fig:HFspec} The Vela pulsar's observed emission spectrum above $100$~MeV as measured by \textit{Fermi} and \textit{H.E.S.S.} \textit{H.E.S.S. I} upper limits for events in the phase intervals of both light curve peaks (P1 and P2) are indicated by the black arrows. Power law (red butterfly) and log parabola (orange butterfly) fits were obtained using \textit{H.E.S.S. II} data. High-energy measurements and an upper limit by \textit{Fermi} are shown as blue points. A power law fit ($>10$~GeV, purple solid line), an exponentially cut off power law fit (cyan dashed line), and a log parabola fit ($>10$~GeV; green solid line) to the \textit{Fermi} data are also indicated \cite{Rudak2015}.}
\end{figure}

Recently, pulsed emission was detected from the Vela pulsar down to 20~GeV with \textit{H.E.S.S.}~\cite{Stegmann2014} and up to 80~GeV with \textit{Fermi} (\cite{Leung2014}, see Figure~\ref{fig:HF_LC}). The second peak in both light curves are in phase. The associated emission spectrum is shown in Figure~\ref{fig:HFspec}. It is probably too early to discriminate between a spectrum that extends to high energies as a power law or cuts off (or breaks) around tens of GeV.

The detection of the Crab pulsar above several hundred GeV prompted \textit{Fermi} to search for pulsed emission from other pulsars at higher energies as well. They detected significant pulsations above 10~GeV from 20 pulsars and above 25~GeV from 12 pulsars \cite{Ackermann2013}. A stacking analysis involving 150 \textit{Fermi}-detected pulsars (excluding the Crab pulsar) was performed by \cite{McCann2014}. However, no emission above 50~GeV was detected, implying that VHE pulsar detections may be rare, given current telescope sensitivities.

Since Geminga is one of the brightest $\gamma$-ray pulsars, it is a promising VHE candidate. Deep upper limits have been obtained by \textit{VERITAS} for this pulsar above 100~GeV \cite{Aliu2015}. Sixty-three hours of \textit{MAGIC} data were analysed by \cite{Bonnefoy2015}, but no VHE pulsed emission was detected. Upper limits for the VHE flux above an energy threshold of 166~GeV have also been obtained by \textit{VERITAS} for the transitional binary pulsar PSR J1023+0038, with 18.1 hours observational time for the radio MSP state and 8.2 hours for the accretion / low-mass X-ray binary state \cite{Richards2015}. Ground-based Cherenkov telescopes as well as water Cherenkov telescopes, e.g., \textit{HAWC} \cite{Alvarez2015}, are now searching for more examples of VHE pulsars.

\section{New theoretical ideas}\label{sec:theories}

All of the standard pulsar emission models (see Section~\ref{sec:history}) predicted HE spectral cutoffs between a few GeV and up to $\sim100$~GeV, generally assuming static $B$-field solutions. Clearly, refinements to these radiation models and $B$-fields are needed to explain the observed VHE emission from the Crab and possibly other bright pulsars. Two new classes of models exist, including inner and outer magnetospheric models.

\begin{figure}[t]
	\centering
	\includegraphics[width=36pc]{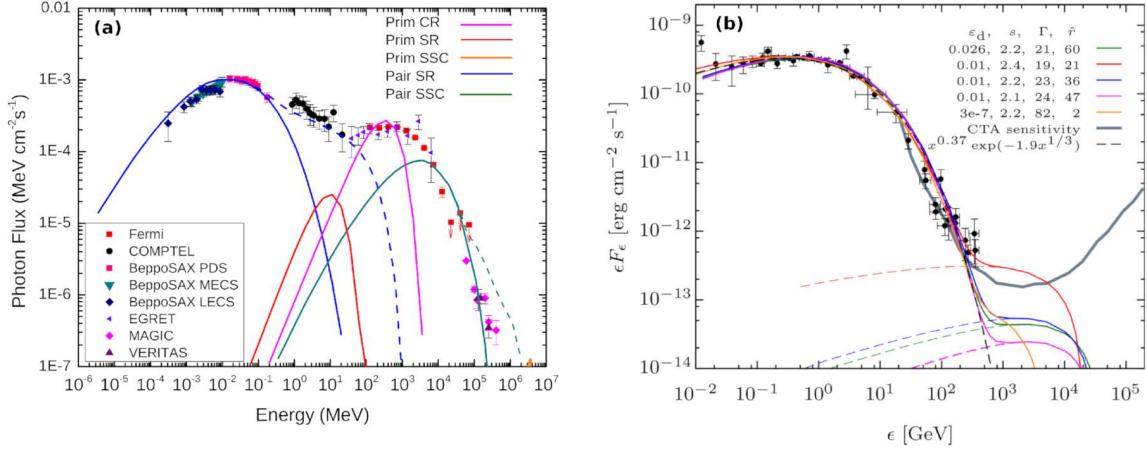}
\caption{\label{fig:SSC} Modelled spectra of pulsed emission from the Crab pulsar. In panel (a) the spectral components from primary electrons and pairs (as labelled) are shown for a magnetic inclination angle $\alpha=45^\circ$, observer angle $\zeta= 60^\circ$ and pair multiplicity $M_{+} = 3\times10^5$ \cite{Harding2015}. In panel (b) the spectrum obtained by \cite{Mochol2015}, for parameters $\epsilon_{\rm d}$, $\hat{r}$ (in units of $R_{\rm LC}$), $s$, and $\Gamma$. They present several best fits for different values of these parameters. The red curve describes the VHE SSC emission best.}
\end{figure}

One new idea is a revised OG model which can produce IC radiation up to $\sim400$~GeV due to secondary and tertiary pairs upscattering infrared to ultraviolet photons \cite{Aleksic2012}. In this OG model, the IC flux depends sensitively on the $B$-field structure near the light cylinder, and the $B$-field lines are assumed to straighten compared to the RVD field. Another idea was proposed by \cite{Lyutikov2012} invoking the SSC radiation process. This is indeed a promising radiation mechanism, but is subject to uncertainties in the injected particle spectral shape and $B$-field structure. The multi-wavelength pulsed emission from the Crab pulsar, from radio-to-TeV, was modelled by \cite{Du2012}, assuming a single-pole annular gap model, although many free parameters are present in this model. The SSC radiation mechanism was applied by \cite{Harding2015} to predict optical to $\gamma$-ray spectra (see Figure~\ref{fig:SSC}(a)), assuming an SG model and a force-free $B$-field structure. This process relies critically on the assumed electrodynamics and the magnetospheric structure. They performed simulations for the Crab and Vela pulsars, as well as two MSPs, i.e., B1821$-$24 and B1937$+$21. However, the only significant predicted SSC component was for the Crab pulsar. For a pulsar to produce significant SSC emission, a high level of non-thermal X-ray emission, i.e., a strong $B$-field at the light cylinder is required so that there will be sufficient soft SR photons for subsequent IC scattering. Also, high pair energies and multiplicity (depending on the $B$-field strength and the period $P$) are necessary for VHE SSC emission. They furthermore tested the addition of an HE power-law extension to the pair spectrum (dashed lines in Figure~\ref{fig:SSC}(a)). The resulting SR spectrum can account for the observed emission in the $1-100$~MeV range. However, the resulting SSC component exceeds the observed \textit{MAGIC} and \textit{VERITAS} points, implying that the observed $1-100$~MeV emission is not produced by the same particles that produce the SSC emission.

There are also models which explain the observed VHE emission from pulsars by invoking radiation from the pulsar wind beyond the light cylinder. For example, \cite{Aharonian2012} modelled an ultra-relativistic wind from the Crab pulsar. This pulsar outflow is dominated by the Poynting flux inside the light cylinder, but abruptly becomes particle-dominated outside (the so-called $\sigma$-problem, where $\sigma$ is the ratio of Poynting flux to kinetic energy flux). They modelled this process by assuming instant or rapid acceleration of particles beyond the light cylinder, with the VHE emission produced at a radius $R_{\rm w}\approx30R_{\rm LC}$ (in units of light cylinder radius $R_{\rm LC}$), with a Lorentz factor $\Gamma_{\rm w}=5\times 10^5$. The predicted light curves only take some geometrical effects into consideration, but not the usual caustic effects, i.e., aberration, retardation, and sweepback of $B$-field lines inside the light cylinder. This should influence estimates of $R_w$ and $\Gamma_{\rm w}$. They predicted VHE emission from IC scattering of the measured power-law X-ray soft-photon field (i.e., the seed photons), and found a sharp energy cutoff at $\sim$500~GeV in the $\gamma$-ray spectrum. Another study proposed a striped wind model \cite{Mochol2015}, using a split monopole $B$-field, in which GeV emission originates in the wavy current sheet. The HE emission arises from SR by relativistic particles that are accelerated via magnetic reconnection. They constrained four parameters, i.e., fraction of magnetic energy dissipated (efficiency) $\epsilon_{\rm d}$, dissipation distance $\hat{r}$ in units of $R_{\rm LC}$, particle spectral index $s$, and Lorentz factor of the wind $\Gamma$, by matching the SR flux to the phase-average spectrum. Various best fits to the spectrum for the Crab pulsar is shown in Figure~\ref{fig:SSC}(b). For the Crab pulsar they found a maximum $\Gamma\approx100$ for low $\epsilon_{\rm d}$ and $\hat{r}$ (orange curve). However, for a larger value of $s$ and $\hat{r}$, and a smaller value of $\Gamma$, the SSC spectral component becomes significantly brighter (red curve), and describes the VHE $\gamma$-ray emission best.

\section{Conclusions and future work}\label{sec:conclusions}

It was not expected that pulsars would be visible in the VHE regime. The detection by \textit{VERITAS} of VHE pulsed emission from the Crab pulsar at energies up to 400~GeV raised important questions regarding the electrodynamics and local environment of pulsars. In view of such observations, refinements to radiation models and $B$-field structures are necessary. More VHE pulsars may be found by the upcoming \textit{Cherenkov Telescope Array (CTA)} which will have a ten-fold increase in sensitivity compared to present-day Cherenkov telescopes. Revised models will assist us in interpreting the VHE emission detected from the Crab and Vela pulsars, and predicting flux levels for other potential VHE pulsars.   

\acknowledgments
This work is supported by the South African National Research Foundation (NRF). AKH acknowledges the support from the NASA Astrophysics Theory Program. CV, and AKH acknowledge support from the {\it Fermi} Guest Investigator Program.

\end{document}